\documentclass[preprint, showpacs ,aps,,amsmath,amssymb]{revtex4}

\usepackage{graphicx}
\usepackage{dcolumn}
\usepackage{bm}

\begin{document}

\title{From supported membranes to tethered vesicles:\\ lipid bilayers destabilisation at the main transition.}

\author{S. Lecuyer}
\author{T. Charitat}
\affiliation{Institut Charles Sadron (UPR 22), CNRS-ULP\footnote{Universit\'e 
Louis Pasteur} 6 rue Boussingault, B.P. 40016, 67083 Strasbourg Cedex France}

\begin{abstract} We report results concerning the destabilisation of supported phospholipid bilayers in a well-defined geometry. When heating up supported phospholipid membranes deposited on highly hydrophilic glass slides from room temperature (i.e. with lipids in the gel phase), unbinding was observed around the main gel to fluid transition temperature of the lipids. It lead to the formation of relatively monodisperse vesicles, of which most remained tethered to the supported bilayer. We interpret these observations in terms of a sharp decrease of the bending rigidity modulus $\kappa$ in the transition region, combined with a weak initial adhesion energy. On the basis of scaling arguments, we show that our experimental findings are consistent with this hypothesis. 

\end{abstract}


\maketitle

\section{Introduction}

Lipid bilayers are very interesting 2D model systems displaying complex behaviours \cite{katsaras, liporevue, lipobudding}; they are also one of the basic building blocks of biological systems \cite{mouritsen}, which explains why they have been the object of growing interest  over the past decades. When studying the physical properties of lipid membranes, two kinds of ideal systems are commonly 
used:  planar membranes supported on hydrophilic substrates, which well-defined 
position enables the use of many experimental techniques, and vesicles, which constitute very simple models for living cells. The most widespread methods for vesicle formation imply the destabilisation of planar lipid bilayers: by an AC electric field in the so-called electroformation method introduced by Angelova et al. \cite{ange}, and also ultrasound \cite{sound}, evaporation \cite{rapid} or freeze drying \cite{freeze}. In spite of its experimental importance, the mechanism of bilayer destablisation has not been elucidated yet. These techniques are mainly empirical, and give limited control of the resulting vesicles. Usually the initial system is a rather disordered lamellar phase, obtained by hydration of a dry lipid film and, unsurprisingly, the result is very polydisperse both in size and lamellarity, making a quantitative comparison with theorical models difficult.

\begin{figure}
\includegraphics[width=7.0cm,angle=0]{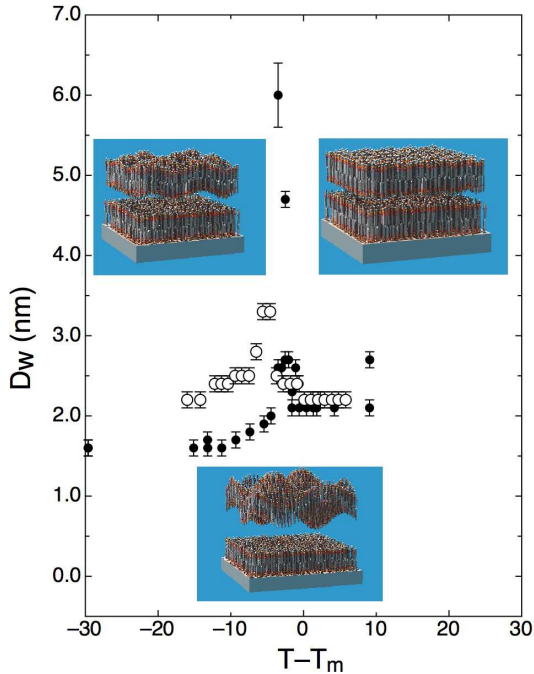}
\caption{Water thickness (D$_w$) separating the two bilayers as a function of temperature for DPPC ($\circ$) and DSPC ($\bullet$) (data from references \cite{fragneto,fragnetoswell}) and schematic representation of the corresponding supported double bilayers.}
\label{f1}
\end{figure}

Double-bilayer systems \cite{charitat1999} give access to a well-localised single "free" membrane. Therefore they enable a local study of the destabilisation of an ideal single bilayer, which makes them promising systems to achieve a better understanding of the mechanism of vesicle formation. Previous neutron and X-ray reflectivity experiments have shown that double-bilayers supported on silicon substrates are stable both in the gel and the fluid phases \cite{fragneto, daillant2005}, but undergo an important swelling around the gel to fluid main transition temperature $T_m$ (fig. \ref{f1}). This was interpreted as a sharp decrease of the bending rigidity modulus $\kappa$ \cite{mecke}.

In this Letter, we show that on different substrates, this swelling can turn into a complete destabilisation. Upon heating single and double bilayers of zwitterionic lipids deposited on glass slides, unbinding occurred in step with the main gel to fluid transition  of the lipids, and lead to the formation of micrometric vesicles of well-defined size. Initially, these vesicles seemed to remain connected to the supported bilayer, but were also able to detach from it in a second time. 

\section{Materials and method} 

Supported single and double-bilayers of DPPC or DSPC (1, 2-dipalmitoyl-sn-glycero-3-phosphocholine and 1, 2-distearoyl-sn-glycero-3-phosphocholine, Avanti Polar Lipids) molecules were prepared using the Langmuir-Blodgett and Langmuir-Schaeffer techniques, as described previously \cite{charitat1999}. The substrates used were glass microscope slides (pre-cleaned, plain, cut edges, 76$\times$26$\times$1 mm, Marienfeld) made highly hydrophilic: sonication for 10 minutes in RBS\texttrademark  \ detergent (Pierce), sonication for 10 minutes in 
ultra-pure water (MilliQ Millipore), then immersion in a fresh "piranha" solution (2/3 
sulfuric acid, 1/3 hydrogen peroxyde) for half an hour. Finally, the slides were thoroughly 
rinsed and sonicated in ultra-pure water for 10 more minutes. Lipids were drawn from a monomolecular layer deposited at the air-water interface of a  Langmuir trough (NIMA Technology) at room temperature, and compressed to high surface pressure (40 mN/m). Starting with the 
hydrophilic substrate immersed in the trough, it was pulled upwards slowly (3 
mm/min), while keeping the surface pressure constant, which resulted in the deposition of a 
first monolayer. A second layer was deposited by lowering the substrate to 
obtain a supported bilayer. To prepare a double-bilayer, a third layer was drawn
similarly, and a fourth one was added using an out-of-equilibrium Langmuir-Schaefer deposition.
In each sample a small fraction ($\sim$ 1\%) of fluorescent lipids was introduced to 
enable fluorescence microscopy observations. The fluorescent probe was 
egg-NBD-PE (L-$\alpha$-phosphatidylethanolamine-N-(4-nitrobenzo-2-oxa-1,3-diazole), 
egg-transphosphatidylated, chicken, Avanti Polar Lipids), a mixture of  headgroup labeled lipids. With this deposition technique, it is possible to select  the labeled layer(s) as needed. 
The quality of the depositions was checked via transfer ratios which were close to 1 for all layers, then by observation of the fluorescence at room temperature; samples were homogeneous, apart from some rare defects probably due to inhomogeneities of the glass slides. As a reference, multilayer systems similar to the ones used for electroformation were also prepared \cite{multi}. Samples were constantly kept immersed in water. 
They were closed while submerged in the trough, using a 5 mm-thick-PTFE spacer and a similar 
glass slide as a lid, and then placed into a water-regulated temperature chamber. The 
temperature was calibrated with a PT100 probe. Observations were made under a microscope (Axioplan 2, Zeiss) equipped with a fluorescence cube and a high-sensitivity CCD 
video camera (Hamamatsu, C2400). Systematic image analysis was carried 
out using the software Visilog (Noesis) \cite{analyse}.

\section{Results}
 
\begin{figure}
\includegraphics[width=8.0cm]{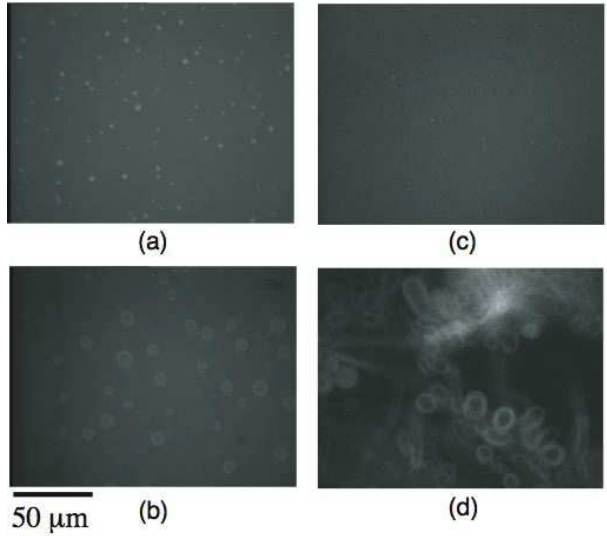}
\caption{Fluorescence microscopy images obtained when heating up: (a) a DPPC bilayer and (b) 
a DPPC double-bilayer deposited on glass substrates; (c) a bilayer deposited on a silicon wafer (almost no vesicles were formed); (d) a thick lamellar phase of DPPC. All images were taken at T=$41.9\pm0.1^\circ$C. }
\label{f2}
\end{figure}

Samples were progressively heated  in steps of $5^{\circ}$C every 30 minutes 
up to $5^{\circ}$C below the main transition temperature ($41.5^{\circ}$C for DPPC, $55.5^{\circ}$C for DSPC \cite{lipidhandbook}). The 
appearance of the sample did not change during this phase. The heating rate was then altered to $1^{\circ}$C every 15 minutes. During this second 
phase, fluorescence inhomogeneities tended to fade from $\sim2^ \circ$C below $T_m$, probably a sign of the increasing lateral mobility of the lipids. At a certain well-defined and very 
reproducible temperature, micrometric vesicles appeared (cf fig.~\ref{f2}). This temperature, we call $T_{ves}$, was always close to the main gel to fluid transition temperature of the lipids $T_{m}$.
For DPPC, $T_{ves}$ 
was found to be between 41 $\pm$ 0.1 $^\circ$C and 41.9 $\pm$ 0.1$^\circ$C. For DSPC, the vesicles 
appeared between 54.2 $\pm$ 0.1$^\circ$C and 55.1 $\pm$ 0.1$^\circ$C.

\begin{figure}
\includegraphics[width=7cm]{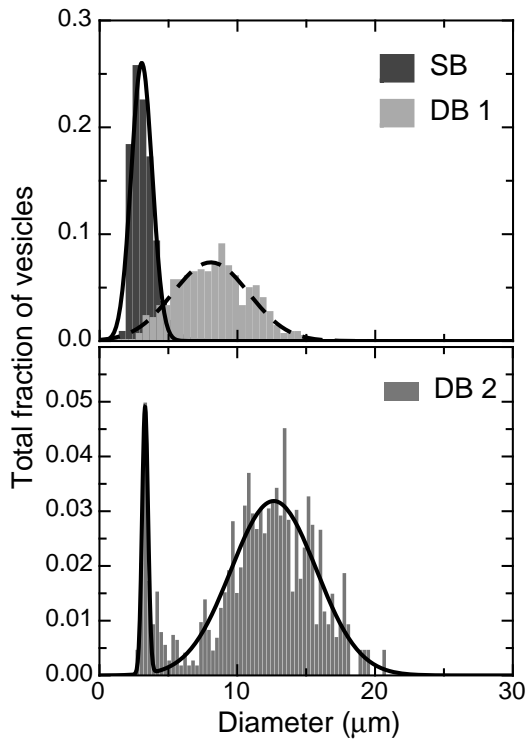}
\caption{Size distribution of the vesicles resulting from systematic  image analysis: (Top) DPPC single
bilayer (SB), diameter = 3 $\pm$ 0.8 $\mu$m (analysis 
of 2516 vesicles) and DPPC double bilayer (DB1),  
diameter = 8 $\pm$ 2.8 $\mu$m (analysis of 
537 vesicles); (Bottom) DPPC double bilayer (DB2),  first peak 
diameter = 3.3 $\pm$ 0.2 $\mu$m , second peak
diameter = 12.6 $\pm$ 3 $\mu$m (analysis of 489 vesicles).}
\label{f3}
\end{figure}

Fig.~\ref{f2} shows typical images observed at $T_{ves}$ for DPPC bilayer, double-bilayer and multilayer systems. With the latter, an irregular multilamellar system, an entire zoology of membranar systems suddenly appears: unilamellar vesicles, onions, tubes...(fig.~\ref{f2} (d)). On the other hand, with well controlled systems such as single and double bilayers, the result is radically different: we observed the formation of rather monodisperse vesicles (see fig.~\ref{f3}). The observed vesicles stand just above the plane of the last bilayer, where they remain for a long time. Initially, their average size increases, which suggests that they 
are still tethered to the membrane. After $\sim$ 20 minutes, their size does not change anymore; a small fraction leaves the surface and can be seen in solution.  After this first stage, no significant change in vesicle size was noticed with time (up to two and a half hours after their formation). 
If the temperature was increased again after vesicle formation, no new vesicles formed - this has been checked up to $\sim 50^\circ$C for a DPPC double bilayer. However further release of the vesicles towards the bulk was favoured, while many of them elongated but could not be detached, confirming the presence of tethers \cite{refere}. If the temperature decreased towards the gel phase, vesicles disappeared from the membrane surface. Either they were all released to the bulk phase, but their low density made them difficult to observe (although a few could be seen), or they were not stable in the gel phase and fused back onto the membrane. This should be elucidated by future experiments. When heating up again, new vesicles were formed.
These observations, along with the reproducibility of the results, strongly corroborates the hypothesis that vesicle formation is a transient phenomenon localised at the main transition. Thus, it is radically different from the thermal unbinding of a lamellar stack at high temperature, as recently described by Salditt et al. \cite{Salditt2000,saldittrevue}.

Figure \ref{f3} shows the size distribution of the vesicles in three cases: a single bilayer with fluorescent probes in the top leaflet (SB); a double bilayer with probes in the second bilayer's top leaflet (DB 1); and a fully labeled double bilayer (DB 2). They exhibit some clear differences:

(i) - Vesicles obtained with a single supported bilayer are significantly smaller (3 $\pm$ 0.8 $\mu$m) than the ones that form from the second bilayer of a double-bilayer (8 $\pm$ 2.8 $\mu$m) which are the only ones we observe here, the fluorescent molecules being localised in the last leaflet.

(ii)  - Size dispersion is slightly higher in the case of the double bilayer, even though gaussian polydispersities are quite similar: 35\% for the double bilayer and 27\% for the single bilayer.
For both types of samples it is important to notice that
polydispersity remains very low compared to typical values for electroformed vesicles 
\cite{spincoat}.

(iii) - A fully labeled DPPC double bilayer (1\% NBD-egg-PE in all leaflets) lead to a 
remarkable bimodal distribution of the size of objects (fig.~\ref{f3}). The 
first peak is in very good agreement with the result of a single bilayer. The second is 
slightly shifted to higher values compared to the distribution found with a double bilayer with 
fluorescent probes in the last leaflet only. This may be related to the large and negatively charged NBD groups, which stand between the two membranes, strongly modifying interaction potentials in the system.

\section{Discussion}

As already mentioned, results derived from neutron reflectivity experiments showed that $\kappa$ collapses a few degrees below $T_{m}$. These experiments also highlighted the stability of single and double bilayers for various neutral phospholipids, in the gel and the fluid phase. Furthermore, no significant change of the reflectivity curves was observed when going through the transition. Typical values 
for the fraction of membrane corresponding to the vesicles seen in our experiments are $\sim 5\%$ 
for bilayers and up to $\sim 20\%$ for double-bilayers, which would significantly modify 
neutron reflectivity curves. We can thus be quite sure that no unbinding occurred in 
neutron experiments. However, a few degrees below $T_{m}$, when $\kappa$ reaches a minimum, an 
important swelling (higher interlayer distance) was seen, along with an increase of the roughness of the second 
bilayer. This different behaviour could be explained by the fact that samples used for neutron reflectivity differ from the ones studied 
here: they contain no fluorescent probe, and the substrates are atomically-flat silicon 
wafers.

Tests were carried out to check the possible influence of the fluorescent label. 
Phase contrast microscopy observations were made difficult by the small size of the vesicles and our 
voluminous thermal chamber, so it was difficult to perform experiments without any fluorescent 
probe. However a DPPC double-bilayer containing only 0.1\% egg-NBD-PE in the 
last leaflet gave no significantly different results.
To test the influence of a possible spontaneous curvature due to the asymmetry of the 
labeled membrane, we performed experiments with a symmetrically labeled DPPC bilayer (1\% probe
in each leaflet), and again the results were very similar. We also carried out neutron reflectivity experiments (data not shown) adding 1\% fluorescent probe to a DPPC bilayer and did not observe any unbinding at the transition.

Thus most probably the relevant parameter is the nature of the substrate. Two parameters can influence the membrane-substrate interaction: the chemical nature of the substrate and its roughness. Here both probably contribute to decreasing the adhesion on glass slides.\\ 
Chemical surface properties of oxidised silicon and glass are quite similar. However one can estimate a reduction of van der Waals interactions by a factor of $\sim$1.5 for a glass substrate compared to silicon coated by a 1-nm-thick oxide layer.
AFM experiments (data not shown) showed no difference for large wavelength roughness (larger than the tip's size $\sim$10 nm) between glass and silicon substrates. However small wavelength roughness could be different: silicon substrates are atomically flat while float glass has a minimum surface roughness (of amplitude $\sim$1-10 nm) due to capillary waves \cite{jackle}. This short wavelength roughness can significantly reduce the local adhesion potential of the supported bilayer \cite{andelman1999, andelman2001}.\\
To check these hypothesis, similar experiments were carried out with a DPPC bilayer supported on a silicon wafer (100, Silchem). Only a few small vesicles formed (see fig.~\ref{f2} (c)), showing that our results are not in contradiction with previous studies. Indeed, in that case the percentage of membrane forming vesicles, $\sim1\%$, stays within the error bars of reflectivity experiments. This is all the more convincing that the wafers used here were not as smooth as the ones specially designed for neutron reflectivity.

Single-component lipid bilayers undergo several changes at the main transition. The area per lipid increases by about 25 $\%$ and the bending modulus strongly decreases due to large density fluctuations \cite{heimburg1, heimburgbook}. Bilayer destabilisation can be induced by both phenomena: mechanical buckling to release excess area and/or thermal unbinding by increase of fluctuations \cite{lipoleible}. Nevertheless, neither vesicle formation nor large length scale defects have been observed on silicon substrates by microscopy or neutron reflectivity, which indicates that supported bilayers can release excess area without dramatic structural modifications. Self-consistent analysis of neutron reflectivity data showed that $\kappa$ decreases from $\sim 200 k_B T$ in the gel phase to $1-3 k_B T$ at the transition, increasing again to $\sim 20 k_B T$ in the fluid phase \cite{mecke}; these results are consistent with macroscopic measurements of $\kappa$ on vesicles \cite{dimova(BiophysJ2000), meleard, mishima} and predictions deduced from heat capacity measurements \cite{heimburgbook}. In classical theory for membrane unbinding \cite{lipoleible, mecke}, bilayer position and fluctuations are controlled by the dimensionless parameter $\beta = \frac{\left( k_B T \right)^2}{A \kappa}$ (A is the Hamacker constant). Unbinding occurs for $\beta_u \sim 0.33$. With the experimental value of $\kappa$, we find that a decrease of the adhesion energy by a factor of 2-3 could induce unbinding at the transition, which is consistent with recent models for rough substrates \cite{andelman1999}. 

In any case, vesicle formation requires large bending of the bilayer. Persistence length of the membrane $\xi_K \equiv \delta\exp{\left[ \frac{4 \pi}{3}\frac{\kappa}{k_BT}\right]}$ ($\delta$ is the molecular length), as introduced by Peliti and Leibler \cite{peliti}, gives the lengthscale beyond which the renormalised bending modulus is smaller than $k_B T$. It requires a value of $\kappa$ around $1.8 k_B T$ to have $\xi_K \sim 5 \mu$m; if one admits vesicle size is correlated to $\xi_K$, this is very consistent with our experimental results. Assuming vesicle formation occurs when the distance between the bilayer and the substrate is comparable to its persistence length, a weaker adhesion energy for the top bilayer will lead to larger vesicles, as we observed experimentally.
However, the actual mechanism is probably more complex and parameters such as the hydrodynamics of the solvent should be considered  \cite{sens}.

\section{Conclusion}

To our knowledge, it is the first time that experimental observation of the unbinding of a single supported phospholipid bilayer is reported. Both {\it mechanical} buckling and thermal unbinding can be involved in membrane destabilisation but an important decrease of the bending rigidity of the membrane in the transition region is required to form vesicles. Low initial adhesion energy due to chemical differences and/or small wavelength roughness could explain why complete unbinding only occurs on glass substrates.\\
These results could give rise to many possible 
developments, such as a new preparation method enabling precise control of vesicle size 
and composition. In particular, vesicle size could be tuned via control of substrate roughness - for instance by etching of silicon wafers.\\
Above all, the results presented here show that supported bilayers are relevant systems for further study of the destabilisation of a single membrane, and a better understanding of the mechanism of vesicle formation.

\acknowledgments
We wish to thank P. Marie for his training on Visilog and his help with image analysis. We thank 
M. Saki, S. Bourbia and N. Schulmann for their participation to this project. We 
have benefited from fruitful discussions with F. Graner, C. Marques, A. Schr\"oder  and  P. Sens.


\begin{thebibliography}{0}
\bibitem{multi}{A few microlitres of 
a lipid solution (10 mg/ml of lipids in chloroform) were spread on a hydrophilic glass slide and left in a 
vacuum chamber overnight. The dry lipid film was then re-hydrated with milliQ water.}
\bibitem{analyse}{Systematic spotting of vesicles was made by  edge detection. Closed boundaries were filled in to make uniform objects; the initial image was turned into a binary picture.
Feret diameters were measured in four directions to check the isotropy of the result. Finally, objects with a form factor larger than 1.2 were not taken into account to avoid any artifact (stuck vesicles, small defects,...).}
\bibitem{refere}{As mentioned by a referee, this could explain the low size dispersion of the vesicles.}
\end{thebibliography}
\end{document}